\documentclass[prl,showpacs,preprintnumbers]{revtex4}
\usepackage{epsfig}
\usepackage{epsf,graphicx,amssymb,amsmath,amsthm}
\usepackage{color}

\definecolor{red}{rgb}{1,0,0}

\begin{document}
\title{Symmetry Induced 4-Wave Capillary Wave Turbulence}
\author{Gustavo D\"uring}
\author{Claudio Falc\'on}
\altaffiliation[Permanent address: ]{Departamento de F\'isica, Facultad de Ciencias F\'isicas y Matem\'aticas, Universidad de Chile.}
\address{Laboratoire de Physique Statistique, Ecole Normale Sup\'erieure,  CNRS, UMR 8550,
24, rue Lhomond, 75005 Paris, France}
\date{\today}

\begin{abstract}
We report theoretical and experimental results on 4-wave
capillary wave turbulence. A system consisting 
of two inmiscible and incompressible fluids of the same density
can be written in a Hamiltonian way for the conjugated pair
$(\eta,\Psi)$. When given the symmetry $z\rightarrow-z$, the set of weakly non-linear interacting waves display a Kolmogorov-Zakharov (KZ) spectrum $n_k\sim k^{-4}$ in wave vector space. 
The wave system was studied experimentally with two inmiscible fluids of almost equal
densities (water and silicon oil) where the capillary surface waves are excited 
by a low frequency random forcing. The power spectral density (PSD) and probability 
density function (PDF) of the local wave amplitude are studied. Both theoretical and experimental 
results are in fairly good agreement with each other.
\end{abstract}


\maketitle

\textit{Introduction:} Wave turbulence\cite{Zakharov} deals with a set of nonlinear random waves in a dispersive medium that, although forced far from thermodynamic equilibrium, can be described statistically. This description is done by means of a kinetic equation for the spectral density distribution $n_k$ which evolves through resonant interactions of $\mathcal{N}$ waves. Besides equilibrium solutions represented by Rayleigh-Jeans distributions, the kinetic equation can display stationary power-law nonequilibrium solutions $n_k\sim k^{-\mu}$, $\mu>0$, called Kolmogorov-Zakharov (KZ) spectra describing the energy exchange (or other conserved quantities) between large and small scales. KZ spectra has been predicted theoretically and observed numerically and experimentally in systems such as bending waves in elastic sheets\cite{placa,Mordant}, Alfv\'en waves in plasma\cite{plasma}, spin waves in solids\cite{spin} and gravity waves in fluids\cite{gravteo,zakgravnum,Fauve1,Denisenko}, to name a few examples. In all the above systems, although the theoretical description depends on several strong constraints (negligable viscosities, density constrasts, aspect ratios, etc\cite{Zakharov,hamiltonian}), experimental and numerical results corroborate the theoretical prediction of the appearence of power-law nonequilibrium spectra. Still, there are certain features that are yet to be studied and compared between theory, experiments and numerics, for instance, the non-gaussianity of the wave amplitudes\cite{Choi}, the nature and existence of intermitency in a wave system\cite{Fauve2,Newell12}, the role of symmetries and dissipation in the wave interactions\cite{Falkovich,Kolmakov} or the deviations of the exponent $\mu$ from the theoretical value.

In this Letter we focus on capillary wave turbulence\cite{capteo,zakcapnum,RestodelMundo} and the effect of symmetries in the wave interactions. We study theoretically and experimentally the statistical properties of random waves at the interface between two inmiscible and incompresible deep fluids of equal densities ($\rho_1$ and $\rho_2$) and depths ($h_1$ and $h_2$). Due to these facts the symmetry $z\rightarrow -z$ is forced on the system: the typical 3-wave capillary wave turbulence breaks down and a four-wave resonant interaction appears as the leading order perturbation. We discuss the effect of this symmetry on the nonlinear type of wave interaction and on the KZ spectrum of the wave amplitude. We compare the theoretical prediction with the experimental measurement of the power spectral density (PSD) of the local wave amplitude at a water-oil interface in the limit of the Atwood number $A=(\rho_1 -\rho_2)/(\rho_1+\rho_2)\rightarrow 0$. Also, the probability density function (PDF) for the excited surface wave amplitude is computed. Both results are contrasted with gravity-capillary wave turbulence measurements.The level of agreement between theoretical and experimental results stresses the fact that capillary wave turbulence is a robust phenomenon for nonlinear random waves.

\textit{Theoretical study:} Let us study the system of potential 
flows of two incompressible and immiscible fluids in a box of height $2h=h_1+h_2$,
where $\eta(r,t)$, $r=(x,y)$ corresponds to the surface elevation between
them, $\rho_1$ is the density of the bottom fluid ($-h_1<z<\eta$), $\rho_2$ the
density of the upper fluid ($\eta<z<h_2$) with $\rho_1>\rho_2$ and $\sigma$ the surface tension coefficient between the two fluids.   
The flows  are defined by the velocity potentials $\phi_1(r,z,t)$ in the lower fluid 
and $\phi_2(r,z,t)$ in the upper fluid with $\nabla\phi_1=v_1$,$\nabla\phi_2=v_2$. It is possible to prove that the dynamics of the two fluid interface have a Hamiltonian structure\cite{hamiltonian,NuevosPeipersdeGustavito}, i.e.,
\begin{equation}
\frac{\partial \eta(r) }{\partial t} =\frac{\delta H}{\delta \Psi}, \ \ \ \ \frac{\partial \Psi(r)}{\partial t} =-\frac{\delta H}{\delta \eta},
\label{hamilton}
\end{equation}
with $\Psi(r)=\rho_1\phi_1(r,\eta(r))-\rho_2\phi_2(r,\eta(r))$ and $H=K+U$ given by
\begin{eqnarray}
H=\int\int^{\eta}_{-h_1}\rho_1\frac{(\nabla \phi_1)^2}{2}dzdr+\int\int^{h_2}_{\eta}\rho_2\frac{(\nabla \phi_2)^2}{2}dzdr\nonumber\\
+\int \left[\frac{g}{2}(\rho_1-\rho_2)\eta(r)^2+\sigma\left(\sqrt{1+\vert\nabla_{\bot} \eta\vert^2}-1\right)\right]dr\nonumber\\
\label{Hamiltonian2}
\end{eqnarray}
where ${\bot}$ correspond to the $r$ coordinates. It is easy to see that the Hamiltonian corresponds precisely to the sum of the kinetic energy $K$ (upper line) and potential energy $U$ (lower line) of the system. The system is also constrained to the bondary conditions $\partial_z\phi_1\vert_{z=-h_1}=\partial_z\phi_2\vert_{z=h_2}=0$ (zero normal velocity at the container walls), $\left[(v_{2\bot}\nabla_{\bot})\eta(r)- v_{2z}\right] _{z=\eta} =\left[(v_{1\bot}\nabla_{\bot})\eta(r)- v_{1z}\right] _{z=\eta}$ (continuity of the normal velocity at the interface) and the incompressibility conditions $\nabla^2\phi_1= \nabla^2\phi_2=0$.
A formal expression can be found for the Hamiltonian \cite{Craig}. The original work was presented for the $2$D case, but can be easily extended for the $3$D case. The kinetic energy can be expressed as 
\begin{equation}
K=\int \Psi \hat G_2(\eta)(\rho_2\hat G_1(\eta)+\rho_1\hat G_2(\eta))^{-1}\hat G_1(\eta) \Psi dr,
\label{KineticNeumann}
\end{equation} 
where $\hat G_1(\eta),\hat G_2(\eta)$ are the Dirichelet-Neumann operators for the fluid domain  $-h_1<z<\eta(r)$ and $\eta(r)<z<h_2$ respectively, defined by $\hat G_i(\eta) \phi_i(r,\eta(r))=(-1)^{i}\left[\nabla_{r}\phi_i\nabla_{r}\eta(r)- \frac{\partial \phi_i}{\partial z}\right] _{z=\eta}$. In 3D systems it does not seem possible to write an explicit Hamiltonian in terms of $\eta$ and $\Psi$. This problem is bypassed by using the "small angle aproximation" to write the Hamiltonian as an infinite Fourier series in $k$-space\cite{zakgravnum,zakcapnum,Zakharov}. In terms of the operators $\hat G_1(\eta),\hat G_2(\eta)$, it corresponds to find an asymptotic series in term of the small parameter $k\eta_0\ll1$, with $k$ the wave vector and $\eta_0$ the characteristic surface elevation. When adding the symmetry $z\rightarrow-z$ to the initial problem (in this case by impossing equal depth and density) the expansion naturally needs to satisfy this constrain. Therefore, the order of the nonlinearity increases from $\mathcal{N}$=3 to 4, and also the system becomes gravity free. In the Hamiltonian expansion $H=H_2+H_4 ...$, one gets 
\begin{eqnarray}
H_2&=&\frac{1}{2}\int \left[\frac{1}{2\rho}k\tanh[kh] \Psi_{\bm k\bm k}\Psi_{-\bm k}+\sigma k^2\eta_{\bm k}\eta_{-\bm k}\right]d{\bm k}\nonumber\\
H_4& =&\int \left( T^{(1)}_{1,2;3,4} \Psi_{{\bm k}_1}\Psi_{{\bm k}_2}\eta_{{\bm k}_3}\eta_{{\bm k}_4} +  T^{(2)}_{1,2;3,4} \eta_{{\bm k}_1}\eta_{{\bm k}_2}\eta_{{\bm k}_3}\eta_{{\bm k}_4} \right) \nonumber\\ 
&\times & \delta({\bm k}_1+{\bm k}_2+{\bm k}_3+{\bm k}_4)d{\bm k}_{1234},
\label{HamiltonianExp}
\end{eqnarray}
where $H_3\equiv 0$,  which comes explicitly from the  the Dirichelet-Neumann operators expansion\cite{Craig}. Physically, the $z\rightarrow -z$ symmetry is not merely imposed by eliminating gravity in the limit $A\rightarrow 0$, but also by impossing equal depth of both fluids, making the capillary surface waves unable to disinguish up from down in the vertical direction. Even more, using the same symmetry arguments, every odd term in the  expansion of $H$ will be zero in this limit. Therefore, the computed transfer matrixes
\begin{eqnarray}
&&T^{(1)}_{1,2;3,4}=\frac{1}{16\rho}(-2\tanh[k_1h]\left(({\bm k}_1\cdot{\bm k}_2)k_1\right)-2\tanh[k_2h]\left(({\bm k}_2\cdot{\bm k}_1)k_2\right)\nonumber\\
&&\frac{1}{16\rho}\left( \frac{[{\bm k}_1\cdot({\bm k}_4+{\bm k}_1)][{\bm k}_2\cdot({\bm k}_4+{\bm k}_1)] }{|{\bm k}_1+{\bm k}_4|}\coth[|{\bm k}_1+{\bm k}_4|h]+\frac{[{\bm k}_2\cdot({\bm k}_4+{\bm k}_2)][{\bm k}_1\cdot({\bm k}_4+{\bm k}_2)] }{|{\bm k}_2+{\bm k}_4|}\coth[|{\bm k}_2+{\bm k}_4|h]\right)\nonumber\\
&&\frac{1}{16\rho}\left( \frac{[{\bm k}_1\cdot({\bm k}_3+{\bm k}_1)][{\bm k}_2\cdot({\bm k}_3+{\bm k}_1)] }{|{\bm k}_1+{\bm k}_3|}\coth[|{\bm k}_1+{\bm k}_3|h]+\frac{[{\bm k}_2\cdot({\bm k}_3+{\bm k}_2)][{\bm k}_1\cdot({\bm k}_3+{\bm k}_2)] }{|{\bm k}_2+{\bm k}_3|}\coth[|{\bm k}_2+{\bm k}_3|h]\right)\nonumber\\
&&T^{(2)}_{1,2;3,4}=-\frac{\sigma}{8}({\bm k}_1\cdot{\bm k}_2)({\bm k}_3\cdot{\bm k}_4),
\label{Tscafinalk}
\end{eqnarray}
depend on four wave vectors. 

Following \cite{Zakharov,placa}, we write $\eta_{\bm k}$=$\frac{X_k}{\sqrt{2}}\sum_sA^s_{\bm k}$ and
$\Psi_{k}$=$-i\frac{X^{-1}_k}{\sqrt{2}}\sum_ssA^s_{\bm k}$, in canonical variables $A^s_{\bm k}$ where $s$=$\pm$ such that $A^+_{\bm k}\equiv A_{\bm k}$, $ A^-_{\bm k}\equiv A^*_{-{\bm k}}$ and $X_k$=$\left(\frac{\tanh[kh]}{2\rho\sigma k}\right)^{(1/4)}$. From the hamiltonian evolution of $A^s_{\bm k}$, a hierarchy of linear equations for the averaged moments ( $\left< A^{s_1}_{ \bm k_1} A^{s_2}_{{\bm k}_2} \right> $, $ \left< A^{s_1}_{ \bm k_1} A^{s_2}_{{\bm k}_2} A^{s_3}_{ \bm k_3} A^{s_4}_{{\bm k}_4}\right> $ and so forth) is written. An assymptotic closure can be given when there exists a separation of linear and non-linear time scales and the system can be regarded as homogeneous in space\cite{Newell12}. The wave spectrum $n_{ \bm k}$, that comes from the second order moment $\left< A_{ \bm k_1} A^{*}_{{\bm k}_2} \right> = n_{ {\bm k}_1} \delta^{(2)}({ \bm k}_1+{ \bm k}_2)$ satisfies a Boltzmann-type kinetic equation describing the slow evolution of the wave spectrum through a four-wave resonant process: 
\begin{eqnarray}
\frac{d}{dt}n_{\bm k}&=&12\pi\epsilon^4 \sum_{s_1,s_2,s_3} \int |L^{-1,s_1,s_2,s_3}_{-{\bm k},{\bm k}_1,{\bm k}_2,{\bm k}_3} |^2  n_{{\bm k}_1} n_{{\bm k}_2} n_{{\bm k}_3} n_{\bm k}\nonumber\\
&\times&\left(\frac{1}{n_{\bm k}}+s_1\frac{1}{n_{{\bm k}_1}}+s_2\frac{1}{n_{{\bm k}_2}}+s_3\frac{1}{n_{{\bm k}_3}}\right)\nonumber\\ 
&\times& \delta(\omega_{\bm k}+s_1\omega_{{\bm k}_1}+s_2\omega_{{\bm k}_2}+s_3\omega_{{\bm k}_3})\nonumber\\
&\times&\delta({\bm k}+{\bm k}_1+{\bm k}_2+{\bm k}_3) d{\bm{k}}_{123}
\label{ecuacioncinetica1}
\end{eqnarray}
where $L^{-s_1,s_2,s_3,s_4}_{-{\bm k}_1 , {\bm k}_2 , {\bm k}_3 , {\bm k}_4}$=$-is_1\frac{1}{24} {\mathcal{P}}_{1234}[ s_1s_2\frac{X_{k_3}}{X_{k_1}}\frac{X_{k_4}}{X_{k_2}}T^{(1)}_{1,2;3,4}-X_{k_1}X_{k_2}X_{k_3}X_{k_4}T^{(2)}_{1,2;3,4}],$ is the scattering matrix, ${\mathcal{P}}_{1234}[\cdot]$ is the sum over the twelve possible permutations of $1,2,3$ \& $4$ and $\epsilon$ is a small parameter related to weak non-linearities which makes possible the time-scale separation. 
With this kinetic equation, we seek isotropic non-equilibrium distribution solutions \cite{zakplasma67}. Despite some differences with the usual kinetic equation, the method of Zakharov can be applied here as in \cite{placa}.
In the deep fluid limit, one finds that the scattering matrix $L$ and frecuency $\omega_{\bm k}=\sqrt{\frac{\sigma}{2\rho}k^3\tanh[kh]}$ of capillary waves are homogenous functions of degree $\beta=3$ and $\alpha=3/2$ respectively, i.e. $L^{s , s_1 , s_2 , s_3}_{\lambda{\bm k},\lambda{\bm k}_1,\lambda{\bm k}_2,\lambda{\bm k}_3}=\lambda^\beta L^{s , s_1 , s_2 , s_3}_{{\bm k} , {\bm k}_1 , {\bm k}_2 , {\bm k}_3}$. 
Looking for a power-law solution of the form $n_k = \Lambda k^{-\mu}$ with $\Lambda$ a constant, it is possible to perform the Zakharov transformation over the right hand side of Eq.(\ref{ecuacioncinetica1}), called collisional terms\cite{zakplasma67}. In such way one gets a stationary out of equilibrium spectrum with $\mu=4$, that represent a constant energy flux solution. If we consider a flux of energy per unit of mass $P$ (it has dimensions of (length/time)$^3$ through the big scales towards the small scales, one can find an explicit expression for $\Lambda$ that leads to $n_k = C P^{1/3} \rho k^{-4}$. $C$ is a pure real number that depends on some integrals directly related with the collisional term that can, in principle, be computed numerically. No inverse cascade is allowed in this system due to the $3\rightarrow1$ wave interaction process, as it was already reported in \cite{placa}. As we are considering a statistically homogenous system in space, it is natural to compute the moments by taking space averages. Nevertheless, from the experimental point of view, taking space averages is quite a difficult task. On the contrary, time averages of local properties are much more accessible. It is possible to relate both for stationary solutions in the linear regime as $ n_T(\omega)d\omega\propto n(k)k^{d-1}dk$, where $n_T(\omega)$ is the time averaged wave number, in frecuency domain. Using $\omega_{ k}$ of capillary waves one gets $n_T(\omega)\propto P^{1/3}\omega^{-7/3}$.  Thus we obtain the spectrum for local surface elevation $\langle\left|\eta_\omega\right|^{2}\rangle_T\propto P^{1/3}\omega^{-8/3}$.

\begin{figure}[h]
\epsfclipon \epsfig{file=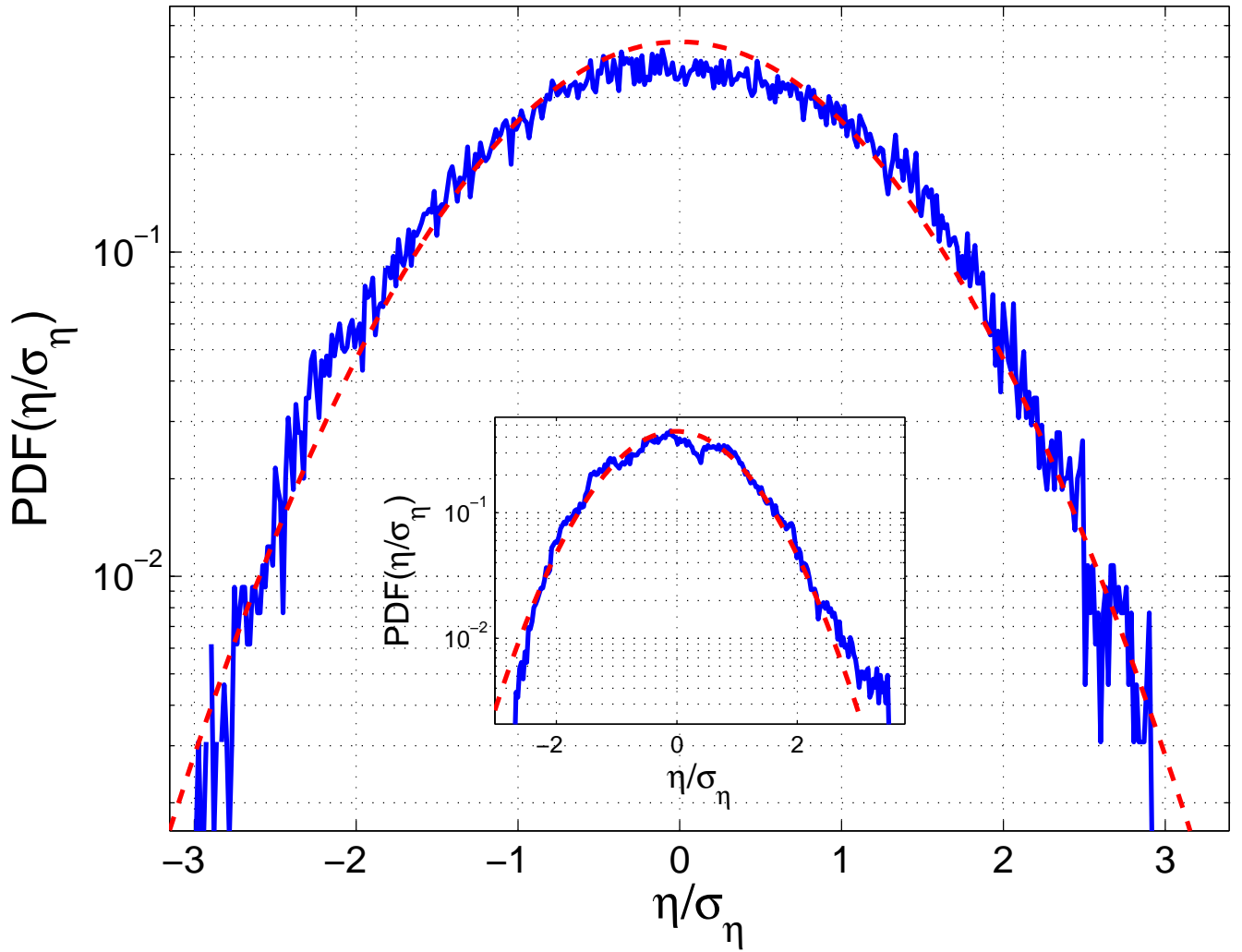,width=\columnwidth}
\caption{(Color online) Probability density function (PDF) of the normalized local wave amplitude $\eta/\sigma_{\eta}$ at the interface of two inmiscible fluids with $A\simeq$0.04 (full blue line) and a parabolic fit (dashed red line). Inset: PDF of $\eta/\sigma_{\eta}$ at the interface of a water-air interface for $A$=1 (full blue line) and a parabolic fit (dashed red line).}
\label{Fig-PDF}
\end{figure}
\textit{Experimental study:} A Plexiglass container (height $h$=60 mm, lenght $l$=100 mm, depth $d$=80 mm) is half filled with destiled water (density $\rho_1=$1.00 g/cm$^{3}$, kinematic viscosity $\nu_1=$0.01cm$^{2}$/s) and half filled with silicon oil (PDMS AB112153 from ABCR, density $\rho_2=$0.93 g/cm$^{3}$, kinematic viscosity $\nu_2=$0.07cm$^{2}$/s). The interfacial tension coefficient $\sigma\sim$ 30 mN/m\cite{Langevin}. The equilibrium interface position is measured at 35 mm. The capillary surface waves are excited by a wave-maker that plunged completely into the upper fluid, oscillating vertically. The wave-maker is driven by an electromagnetic vibration exciter via a power amplifier. The random forcing, supplied by the source output of a dynamical spectrum analyser, is low-pass filtered between 0-$f_{driv}$=3 Hz. The excited surface wave amplitude $\eta$ is locally measured 4 cm away from the container walls by means of a wire capacitive gauge, 0.1 mm in diameter. The measured capacitive fluctuations are proportional to the local wave amplitude ones.  They are sampled at 800 Hz during 300 s, and filtered numerically at 500 Hz to avoid aliasing. The only noteworthy difference in this set-up that in this case both the dielectrics are liquids of similar densities and similar viscosities. We have checked the linear response in the local wave amplitude $\eta$ by changing fluid depths and the constant frequency response (in magnitude) of the wire probe for the water-oil boundary in a frequency band between 1 to 100 Hz.

\begin{figure}[h]
\epsfclipon \epsfig{file=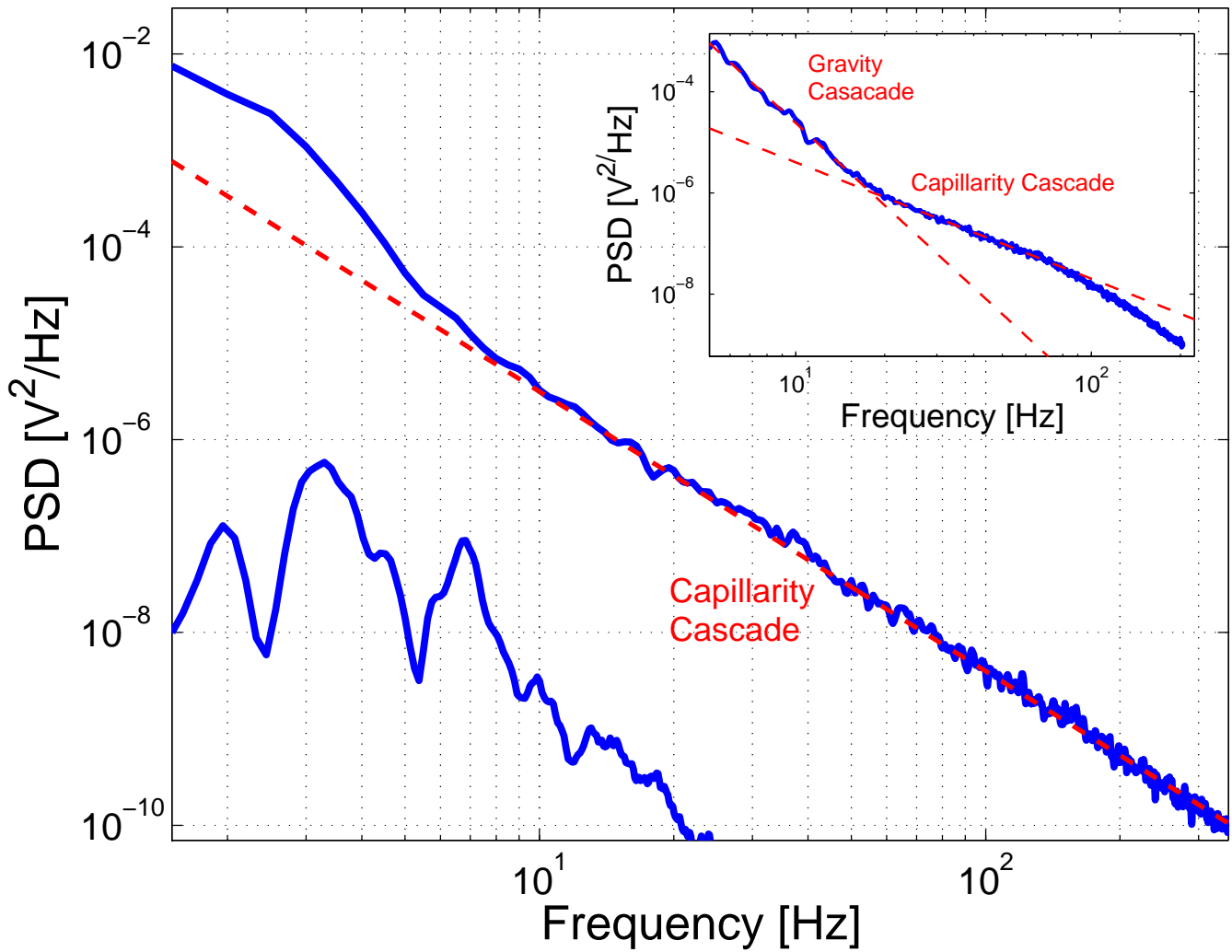,width=\columnwidth}
\caption{(Color online) Power spectral densities (PSD) of the local wave amplitude $\eta$ at the interface of two inmiscible fluids with $A\simeq$0.04 (full blue line) for low (bottom) and high (top) forcing amplitudes. Best fit slope -2.75 (dashed red line). Inset: PSD of $\eta$ at a water-air interface for $A$=1 (full blue line) and best fit KZ spectra (dashed red line) for gravity (-5.35) and capillary (-2.52) waves.}
\label{Fig-PSD}
\end{figure}

With the acquired data, the probability distribution function (PDF) of $\eta$, normalized by its {\it rms} fluctuations $\sigma_{\eta}$ is calculated, as shown in Fig. \ref{Fig-PDF} (main). Notice that $\left<\eta\right>\sim0$ and that its fluctuations are close to being symmetric with respect to $\eta=0$. No exponential tails are found. The kurtosis is slightly larger than 3, but not large enough to exclude gaussianity. For comparison, we show in Fig. \ref{Fig-PDF} (inset) the PDF of $\eta/\sigma_{\eta}$ when gravity-capillary wave turbulence developes. We see a clear asymmetric tail (positive skweness) as it has been shown elsewhere\cite{Fauve1}. This contrast is a clear indication of the symmetry imposed in the system: there is no external field (such as gravity) that breaks the $z\rightarrow-z$ parity so the surface perturbations are symmetric with respect to $\eta=0$. It is unclear if the fluctuations are indeed gaussian, but resolution of large events could not be made in the present experimental set-up.
The wave system has a very low Atwood number $A=(\rho_1-\rho_2)/(\rho_1+\rho_2)\simeq$0.04, reducing the effective gravity drastically. One finds that in this system the capillary length $l_c=2\pi\sqrt{A\sigma/g(\rho_1+\rho_2)}$ where the crossover from gravity to capillary regime takes place is an order of magnitude larger than in a liquid-air interface problem\cite{Cazabat}. The frequency crossover between gravity and capillary regimes $f_c=\omega_c/2\pi=\pi \sqrt{ Ag / 2 l_c}$ will be obtained at a frequency close to 3-4 Hz. Therefore, when the frequency cut-off of the forcing is larger than $f_c$, the only KZ-type spectrum we can observe is the capillary one. In Fig. \ref{Fig-PSD} we show both the pure capillary (main figure) and the gravity-capillary (inset) powerlike spectra. For pure capillary waves, as the forcing amplitude is increased, low frequency normal modes and harmonics of $f_{driv}$ dissapear and a power-law spectrum developes. Only one scale-invariant spectrum with slope $\sim -2.75\pm0.05$ appears in the capillary-driven transparency window (for frequencies larger than the characteristic frequencies of the broad-band forcing), which is within the experimental error from the theoretical $f^{-8/3}$. In this wave turbulent regime, no cusps over the wave crests were observed, which sustains the assumption $k\eta_0\ll$1 and eliminates the possibility of singularities polluting the spectral content of the signal. The gravity-capillary spectra of Fig. \ref{Fig-PSD}(inset) is calculated from the local elevation of surface waves when the lighter fluid (PDMS) is removed. Waves are excited using a wavemaker pluging in water driven by a low-frequency ($f_{driv}$=3 Hz) random forcing similar to \cite{Fauve1,Fauve2}. Both gravity $f^{-5.35}$ and capillary $f^{-2.52}$ wave spectra are within the experimental range of \cite{Fauve1}.

\textit{Conclusions}: In this paper we show the appearence of 4-wave capillary turbulence at the interface of two inmiscible and incompresible fluids with almost equal densities, due to an imposed spatial simmetry. The theoretical wave spectrum of the amplitude fluctuations behaves as $f^{-8/3}$ in frequency domain and $k^{-4}$ in wave vector space. Experimentally, we have computed the PDF and PSD of the local amplitude fluctuations $\eta$ at the interface of an oil-water mixture. The computed PDF of $\eta$ has a gaussian form, and no exponential tails where found\cite{Fauve1}. The experimental PSD of $\eta$ shows a power-law behavior $\sim f^{-2.75}$ in agreement with the expected theoretical slope for 4-wave interaction process. The excited surface waves in the experimental set-up display wave turbulence-type behavior which agrees with the theoretical prediction, even though $A\neq$0 and both viscosities are not small, as the theory necesitates. A point worth studying both experimentally and theoretically is their effect on the dissipation scale and the effect of different depths, specially in the limit $h_1/h_2\ll1$. It must be noticed that the computed spectrum is close to the theoretical $f^{-17/6}$ for 3-wave interaction processes. This slight departure can be explained by the fact that the 3-wave process may not suppressed completely due to a slight asymmetry in densities. The study of the slope dependence on $A$ will be presented elsewhere.

The authors would like to acknowledge the support of Asefe and CONICYT and fruitful discussions with S. Fauve, F. Petr\'elis, E. Falcon, S. Rica and A. M. Cazabat.


\begin{thebibliography}{let25}

\bibitem{Zakharov} V. E. Zakharov, V. S. L'vov, and G. Falkovich, \textit{Kolmogorov
Spectra of Turbulence I} (Springer, Berlin, 1992).

\bibitem{placa} Gustavo D\"uring, Christophe Josserand and Sergio Rica,  Phys. Rev. Lett. {\bf 97}, 025503 (2006)

\bibitem{Mordant}N. Mordant, Phys. Rev. Lett. {\bf 100}, 234505 (2008); A. Boudaoud, O. Cadot, B. Odille and C. Touz\'e, Phys. Rev. Lett.{\bf  100}, 234504 (2008)

\bibitem{plasma} R. Z. Sagdeev, Rev. Mod. Phys. {\bf 51}, 1 (1979)

\bibitem{spin}  V. S. L'vov, {\it Wave Turbulence Under Parametric Excitation} (Springer-Verlag Berlin, 1994)

\bibitem{gravteo} K. Hasselmann, J. Fluid Mech. {\bf 12}, 481 (1962); 15, 273 (1963); V. E. Zakharov, N. N. Filonenko, Dokl. Akad. Nauk. SSSR {\bf 170} (6) (1966) 1292–1295 

\bibitem{zakgravnum}A. N. Pushkarev and V. E. Zakharov, Phys. Rev. Lett. {\bf 76}, 3320 (1996).

\bibitem{Fauve1} E. Falcon, C. Laroche and S. Fauve, Phys. Rev. Lett., {\bf 98}, 094503 (2007)

\bibitem{Denisenko} Petr Denissenko, Sergei Lukaschuk and Sergey Nazarenko, Phys. Rev. Lett. {\bf 99}, 014501 (2007)



\bibitem{capteo} V. E. Zakharov, N. N. Filonenko,  Zh. Prikl. Mekh. I Tekn. Fiz. {\bf 5}, 62–67 (1967)

\bibitem{zakcapnum}  A. I. Dyachenko, A. O. Korotkevich and V. E. Zakharov, Phys.Rev. Lett. {\bf 92} 134501(2004); A. N. Pushkarev and V. E. Zakharov, Phys. Rev. Lett. {\bf 76}, 3320 (1996) 

\bibitem{RestodelMundo} M. Yu. Brazhnikov, G. V. Kolmakov, A. A. Levchenko and L. P. Mezhov-Deglin, Europhys. Lett., {\bf 58} (4),  510–516 (2002); William B. Wright, Raffi Budakian, and Seth J. Putterman, Phys. Rev. Lett. {\bf 76}, 4528 (1996);  E. Henry, P. Alstrøm and M. T. Levinsen, Europhys. Lett., {\bf 52} (1), 27–32 (2000); R. Glynn Holt and Eugene H. Trinh, Phys. Rev. Lett. {\bf 77}, 1274 (1996)

\bibitem{hamiltonian} V. E. Zakharov (Editor), \textit{Nonlinear Waves and Weak Turbulence}  (American Mathematical Society Translations Series 2 Volume 182, 1998)

\bibitem{Choi} Yeontaek Choi, Yuri V. L'vov, Sergei Nazarenko, Physica D {\bf 201}, 121 (2005)

\bibitem{Fauve2} E. Falcon, C. Laroche and S.Fauve, Phys. Rev. Lett., {\bf 98}, 154501 (2007)

\bibitem{Newell12} A. Newell, S. Nazarrenko and L. Biven, Physica D (Amsterdam), {\bf 152-153}, 520 (2001)

\bibitem{Falkovich} I.V. Ryzhenkova and G.E. Falkovich, JETP {\bf 71}, 1085 (1990)

\bibitem{Kolmakov} G.V. Kolmakov, JETP {\bf 83}, 58 (2006)

\bibitem{NuevosPeipersdeGustavito} T. B. Benjamin and T. J. Bridges, J. Fluid Mech. {\bf 333}, 301-325 (1997)

\bibitem{Craig}  W. Craig, P. Guyenne and H. Kalisch, Comm. Pure Appl. Math., {\bf 58}, 1587–1641 (2005)

\bibitem{zakplasma67} V. E. Zakharov, Zh. Eksper. Teoret. Fiz. {\bf 51}, 686 (1966) 

\bibitem{Langevin} V. Bergeron and D. Langevin, Phys. Rev. Lett. {\bf 76}, 3512 (1995)

\bibitem{Cazabat} Although $\sigma$ depends on both water and PDMS chemical origins and treatments, at this value of viscosities, it will always remain larger than 10 mN/m, making $f_c$ an order of magnitude smaller than in the typical fluid-air problem (A. M. Cazabat, private comunication) 


\end{thebibliography}
\end{document}